\documentclass[12pt]{article}
\usepackage{amsmath}
\usepackage{setspace}
\usepackage{rotating}
\usepackage{graphicx}
\usepackage{varioref}

\begin{document}

\title{Variations in Sonoluminescence Flash Timing}

\author{Thomas E.\ Brennan \\ \emph{West Virginia Wesleyan College} \\ 59 College Dr.\ Box 126 \\ Buckhannon, WV 26201}


\date{\today}

\maketitle

\begin{abstract}

Since the first experimental results were published in the 1990s, it has been believed that the sonoluminescence flash always occurs no more than a few nanoseconds before the minimum radius of a collapsing bubble.  A concurrent belief has been that the period between sonoluminescence flashes is steady on the order of a few nanoseconds, and that sonoluminescence flashes occur with a ``clock-like'' regularity.  To the contrary, data presented here show that the sonoluminescence flash can occur hundreds of nanoseconds before the minimum radius and that the sonoluminescence flash-to-flash period can vary on the order of hundreds of nanoseconds.  These new findings may require a reexamination of the physics of sonoluminescence. 

\end{abstract}


\section{Introduction}

Sonoluminescence is the emission of light flashes by sonically oscillating gas bubbles in a fluid.  Sonoluminescence is brighter if a noble gas has been dissolved into the fluid -- typically a water-glycerine mixture which has been degassed and then regassed with either argon, krypton or xenon.  

Our work was limited to an examination of \emph{single-bubble sonoluminescence}, SBSL, because it is stable, and hence easier to observe and measure than \emph{multi-bubble sonoluminescence}, MBSL -- although we expect that what we have learned applies to MBSL as well.

In SBSL, the sonoluminescent flash occurs only once per acoustic period, is very brief, and is very close to the moment of minimum bubble radius in the cycle of bubble expansion and contraction.  Exactly how close to the time of minimum radius, and the degree to which that timing can vary, is the object of this investigation.  We were motivated to repeat the timing studies done by others \cite{PuttermanMieScattering}, \cite{GaitanThesis}, not because we believed their results to be inconclusive, or unclear, but because we suspected that the observation of pulses with highly-regular timing might be an exceptional case resulting from a very well-controlled, smooth resonance, or some other conditions.  We suspected that it might be possible to observe sonoluminescence in which the timing of the pulses was variable or farther away from the minimum radius.  In fact, other published studies have hinted at this \cite{CarlosMegahertz}.

We have discovered that for the 23.55 kHz system which we constructed, the SL flashes can occur hundreds of nanoseconds before the time of minimum bubble radius.  At those times, the fluid-dynamic models imply that the contents of the bubble should be close to room temperature, and that the motion of the bubble wall should be subsonic -- far too slow to have created a shock-wave plasma, and well before any compressive heating could have occured.

\section{Experimental Design}


To observe the motion of the bubble and the timing of the flash in the bubble cycle, we designed a Mie-scattering experiment (Figure \vref{laserScatteringOverhead}), similar to that used by other authors, \cite{PuttermanMieScattering}, \cite{CarlosMegahertz}.  

The procedure begins by preparing a mixture of water and 17\% glycerine by volume, which has been degased and then brought into equilibrium with xenon, krypton or argon at 1/3\,rd atmosphere.  This solution is then poured into a 200 mL spherical glass flask with a narrow neck which has a resonant frequency around 23.55 kHz.

A frequency generator, audio amplifier and impedance-matching inductor are used to drive PZTs attached to the sides of the flask, which is mounted on a labstand.  When the frequency and drive are appropriate, bubbles can be trapped and made to glow by disturbing the surface of the water with a syringe. 

To begin the light scattering measurements, a 532 nm continuous-wave laser is aimed at the glowing bubble.  The SL light and scattered laser light from the bubble are then focused onto a shielded Hamamatsu PMT, model R329-01, with a 3-inch condenser lens.  The signal from the scope was recorded with a Tektronix 7104 digital sampling oscilloscope with a bandwidth of 3 GHz and a maximum sampling rate of $10 \text{\,GS/s}$.  Our 2008 data were recorded with a 300 MHz Tektronix TDS 3034 sampling oscilloscope with a maximum sampling rate of $3 \text{\,GS/s}$.  This setup can be used to observe only the SL flash by turning off the laser.

\section{Results}

\subsection{Flash-to-Flash Jitter}Before we ask how the timing of the flash varies with respect to the minimum bubble radius,  let us first investigate how the timing between SL flashes varies.  We call this the flash-to-flash jitter. 

We took these data with the Tektronix 7104 oscilloscope in November 2009. We recorded twenty scope traces of five sequential SL flashes like those shown in Figures \ref{KryptonFlashToFlash3} and \ref{KryptonFlashToFlashZero}.  Thus we have a dataset of 100 flashes, from which we are be able to calculate 80 datapoints of flash-to-flash jitter.  These were all taken with a resolution of 800 ps per sample.  The flash times are identified with a simple algorithm which picks out the sample time corresponding to the maximum of the peak, (most negative voltage).     

Tables \ref{KryptonFlashToFlashTable3} and \ref{KryptonFlashToFlashTableZero} summarize the flash data from Figures \ref{KryptonFlashToFlash3} and \ref{KryptonFlashToFlashZero}.  We have presented the average period for the five flashes, the average frequency, the time of each of the five flashes, and the jitter, which is the amount which the four sequential flash-to-flash times differ from the average flash-to-flash time.  The flash-to-flash jitter data from all 20 traces was calculated in this way and is presented as a histogram in Figure \ref{KryptonFlashToFlashHistogram}.  Figure \ref{KryptonFlashToFlashA} shows a zoom-in on the first flash from Figure \ref{KryptonFlashToFlash3}.  (Note, due to constraints of space, we were not able to provide zoom-ins on each of the flashes from Figure\ref{KryptonFlashToFlash3}, we will include these at a future time if we receive editorial permission.)


Figure \ref{KryptonFlashToFlash3} and Table \ref{KryptonFlashToFlashTable3} show an exceptional case where the flash-to-flash jitter was as large as 102 ns.   
 
In   Figures \ref{KryptonFlashToFlash3} and \ref{KryptonFlashToFlashZero} an oscillatory noise pattern with the same period as the SL is visible on the scope traces.  It is more prominent in Figure \ref{KryptonFlashToFlash3}.  This noise is RF interference from the audio amplifier and cabling which induced a small voltage in the PMT signal.  This noise signal has a gentle slope compared to sharpness of the SL flash, and should not affect our measurement of the peak SL time.  In addition, the data in Figure \ref{KryptonFlashToFlash3} and \ref{KryptonFlashToFlashZero} have a noise floor.  (It is more prominent in Figure \ref{KryptonFlashToFlash3}).  However, the SL signal is discernable above this noise floor, and because of the narrowness of the SL flash, this noise cannot move the peak SL time by more than 5 ns.  

Now, looking at Figure  \ref{KryptonFlashToFlashA}, we see that the SL flash has a \emph{full-width half-maximum} of 5 ns, so that at most 10 ns of the jitter might be due to noise or the irregular response of the PMT to the flash.  However, the observations of 20 to 100 ns jitter as shown in the histogram in Figure \ref{KryptonFlashToFlashHistogram} must be real and not due to systematic error.  It has been previously reported that the SL flash has a ``clock-like'' precision, with a jitter on the order of picoseconds \cite{PuttermanReview}.  

Also note from examining Figures \ref{KryptonFlashToFlash3} and \ref{KryptonFlashToFlashZero} how the intensity each flash can vary.

\subsection{Flash-to-Minimum-Radius Timing}

Here we present our measurements of the timing of the sonoluminescent flash with respect to the minimum bubble radius.  We performed this experiment in January and February 2008  and then again in November 2009.  We present here a just a few of the hundreds of recorded flash events from sonoluminescing argon, krypton and xenon bubbles.  We believe that to provide the most concise presentation and to eliminate the possibility of a phase offset from different signal paths, the laser scattering and SL flash should be observed with a single PMT.  

Figures \ref{fig:091120012ManualFit} to \ref{lateXenon} depict scope traces showing both the laser scattering and SL flash, and give an indication of our measurement of the time difference between the occurrence of the flash and the time of minimum bubble radius.  This time difference can clearly be seen by eye, but to improve the likelihood that our estimate of the time difference is accurate we have also fit these scope traces to solutions of the Rayleigh-Plesset-Keller equation.   

Let us present a brief discussion of this equation and how its solutions can be fit to our raw laser scattering data. 

The oscillation of a gas bubble in a fluid can be described by the Rayleigh-Plesset-Keller equation:
\begin{eqnarray}
& & \hspace{-20mm} \left(1-\frac{\dot R}{v_s}\right) \rho R \ddot R + \frac{3}{2} \rho {\dot R}^2 \left(1-\frac{\dot R}{3 v_s}\right)  \nonumber \\
&=&\left(1+\frac{\dot R}{v_s}\right)(P_{\text{gas}}-P_0-P(t)) +\frac{R}{v_s} \dot P_{\text{gas}} - 4 \eta \frac{\dot R}{R} - \frac{2 \sigma}{R}
\label{RPKEquation}
\end{eqnarray}
Here, $R(t)$ is the bubble radius as a function of time, $v_s$ the speed of sound in the fluid, $ \eta$ the dynamic viscosity of the fluid, $\sigma$ the surface tension, $P_0$ the ambient pressure, $P(t)$ the pressure of the acoustic drive, and $P_{gas}$ the pressure of the gas in the bubble.  

This differential equation must be supplemented with an expression for $P_{gas}$ as a function of bubble radius:
\begin{equation}
P_{\text{gas}} (R) = P_{\text{gas}}(V(R)) = \frac{P_0 {V_0}^{\gamma}}{(V - n b)^\gamma}
\label{RPAdiabatic}
\end{equation}
Here, $b$ is the van der Waals excluded volume per mole. Such a formulation also requires that the temperature of the gas in the bubble varies as:
\begin{equation}
T_{\text{gas}} (R) = T_{\text{gas}}(V(R)) = \frac{T_0 {V_0}^{ \gamma - 1 }}{(V - n b)^{ \gamma - 1 }}
\label{RPTemp}
\end{equation}
Note that a radial expansion or compression factor of ten will result in a hundred-fold decrease or increase in temperature, $\gamma$ being $5/3$ for a monatomic gas such as xenon. 

The solution to the RPK equation (\ref{RPKEquation}) is characterized by two experimental parameters -- the equilibrium radius of the bubble $R_0$ and the amplitude of the drive, $P_d$.  $R_0$ is the radius which the bubble would have if there were no drive present, which we can consider as the size of the injected bubble.  When the bubble passes through or returns to this radius during its adiabatic cycle, it is at atmospheric pressure and room temperature.  Since the RPK equation is second order in time, its solution also requires the specification of the initial velocity of the bubble wall, $ \dot{R}(t=0) $.  However, these various solutions all settle into the same phase and response with respect to the drive after one or two periods have elapsed.   So once the two parameters $R_0$ and $P_d$ are specified, a unique solution to the RPK equation, up to an overall time offset, can be generated.

To match solutions of the RPK equation to light scattering curves requires an additional hypothesis on how the bubble scatters light and how the PMT responds to that light.  Other researchers  \cite{PuttermanMieScattering} have had success with the the hypothesis that the intensity of scattered light, and hence PMT voltage, is proportional to the cross sectional area of the bubble.  Thus,
\begin{equation}
V(t) = \alpha R^2 (t+t_{\text{offset}}) + V_{\text{bkg}}
\label{MieTheory}
\end{equation}
where $V_{\text{bkg}}$ is the amplitude of the laser light (PMT voltage) when no bubble is present.  In order to calibrate the data we must find a value of $\alpha$ and $V_{\text{bkg}}$.  Our primary goal in fitting the data is to obtain a value of $t_{\text{offset}}$, so that the time of minimum bubble radius can be determined from the fit.

With this simple light scattering model, five parameters are required to fit voltage data to the solution $R(t)$ of the RPK equation: $R_0$, $P_d$, the time offset $ t_{\text{offset}}$, the factor $\alpha$ and the background voltage $V_{\text{bkg}}$.    Hence there will be a five parameter space in which to minimize the \emph{sum of squared differences}.  This can be challenging and unreliable, as the space may be filled with spurious local minima.  Informed initial guesses of the parameters to be varied will improve the likelihood of good convergence.  

In practice, for our 23.55 kHz system,  we have found that the values of $R_0$ and $P_d$ do not vary greatly and are typically around $ 9 \; \mu m $ and $ 1.5 \; atm $ respectively.  Knowing this, we can fix $R_0$ and $P_d$, and make the fitting procedure less computationally intensive.  We performed all of these fits using the \emph{NonlinearModelFit} routine in \emph{Mathematica v.\  7.0}.

The data in Figure \ref{fig:091120012ManualFit}  depict two entire oscillation cycles.  We can see the bubble grow, and then bounce several times before the cycle repeats.  This data is fit to a solution of the RPK equation with $R_0 = 8.7 \, \mu m$ and $P_d = 1.62 \, atm $.  At this scale the SL flashes are barely visible; but the next two Figures, \ref{091120012aZFit} and \ref{091120012bZFit}, show zoom-ins on the first and second flashes, with the flash clearly visible before the minimum radius.  In these two figures the fit was recalculated only allowing $\alpha$, $t_{\text{offset}}$, and $V_{\text{bkg}}$ to vary to improve alignment  with the minimum radius using a 2000 ns window of data.  

Figures \ref{typicalArFlashB} thru \ref{lateXenon} depict zoom-ins near the minimum radius from the 2008 data set.  These have been fit to solutions of the RPK equation with with $R_0$ fixed at $9.5\,\mu m$, $P_d$ fixed at $1.4\,\text{atm}$.  Figure \ref{lateXenon} show a case where the peak of the flash is apparently 60 ns \emph{after} the minimum radius.  A close inspection of this figure suggests that the flash may have begun right at, or just a few nanoseconds after the minimum radius.  If this is not truly a late flash, then at least it underscores the variability in flash timing when compared to the other traces.

An added bonus of performing these fits is that we are able to obtain a calibration of the laser scattering data, and hence determine the velocity of the bubble wall at the moment of SL.  Table \ref{table:FlashToMinRadius}  summarizes the timing and bubble wall velocity measurements from Figures \ref{fig:091120012ManualFit} to \ref{lateXenon}.

\section{Conclusion}

The assertions that sonoluminescence has a ``clock-like'' regularity, or that the flash \emph{always} occurs within a few nanoseconds of the minimum radius can be disproven simply by  observing a single counterexample.  We have presented several direct counterexamples to those assertions.

We have attempted to depict our data in a manner which makes it obvious to the eye that the sonoluminescence flash timing is variable.    It may be possible that a careful reviewer may find fault or take exception with some aspect of our method of fitting the Rayleigh-Plesset-Keller solutions to the light scattering data.  Ideally, in a longer article, we would have room to go into more detail about our fitting methods.  But we believe that the fits at least provide a corroboration of what is already quite apparent to the eye.  If our calibration of the curves is accurate, then we have also shown the that bubble wall motion is subsonic at the moment of SL.

It is obvious from our data that the sonoluminescence flash timing can vary on a scale much larger than previously reported.  The sonoluminescent flash can occur hundreds of nanoseconds before the minimum radius, at times for which the bubble wall motion is subsonic, so that it seems that the flash cannot be the result of a thermal shock plasma formed by the collapse of the bubble.  \emph{Thermal-shock} models logically require that the flash occur right at or slightly before the minimum radius.  For example, a shockwave model proposed by Greenspan \cite{Greenspan} and Wu \cite{WuSimulation} requires the SL flash to occur just one tenth of a nanosecond before the minimum radius, because it is not until that moment in time that the RPK model implies the bubble wall velocity is large enough to create a thermal shock.  Thus, that model and others like it must be ruled out.  Furthermore, those models assume that the RPK model is an accurate description of bubble motion at those times, which may not be the case -- those models do not take into account the \emph{boiling} of the bubble wall that might dissipate the energy of the collapse before a thermal shock could be created.  

In a follow-up to this article we will introduce a new model of sonoluminescence which proposes that the SL flash is emitted by an excited cold condensate formed during the adiabatic expansive cooling which breaks apart to emit light and energetic electrons at random times near the minimum radius.



\begin{table}[htp]
\begin{center}	
\caption{\label{KryptonFlashToFlashTable3}  A table of the five flash times from Figure \ref{KryptonFlashToFlash3}.}
\begin{tabular}{  c  c  c  c  c }
\\
\hline \hline
Flash & Avg. Period (ns) & Frequency (Hz) & Flash Time (ns) & Jitter (ns)   \\ 
\hline 
A     &  --              &       --         &  8970.4         &     --        \\ 
B     &  --              &       --         &  51405.6        &    -27.    \\ 
C     &  42462.2         &    23550.4       &  93946.4        &    78.6    \\ 
D     &  --              &       --         &  136306.        &    -102.2    \\ 
E     &  --              &       --         &  178819.        &    50.6    \\ [1ex]
\hline
\end{tabular} 
\end{center}   
\end{table}

\begin{table}[htp]
\begin{center}	
\caption{\label{KryptonFlashToFlashTableZero}  A table of the five flash times from Figure \ref{KryptonFlashToFlashZero}.}
\centering
\begin{tabular}{  c  c  c  c  c }
\\
\hline \hline
Flash & Avg. Period (ns) & Frequency (Hz) & Flash Time (ns) & Jitter (ns)   \\ 
\hline 
A     &  --              &       --         &  14980.8         &     --        \\ 
B     &  --              &       --         &  57476.8        &    -13.2    \\ 
C     &  42509.2         &    23524.3       &  99991.2        &    5.2    \\ 
D     &  --              &       --         &  142498.        &    -2.0    \\ 
E     &  --              &       --         &  185018.        &    10.0    \\ [1ex]
\hline
\end{tabular}
\end{center}    
\end{table}

\begin{table}[htp]
\begin{center}	
\caption{\label{table:FlashToMinRadius} A table of the times of the flash with respect to the minimum radius, $\Delta t$  and bubble wall velocities from Figures \ref{091120012aZFit} thru \ref{lateXenon}.  Positive $\Delta t $ values indicate time \emph{before} the minimum radius, negative values, time \emph{after} the minimum radius. }
\begin{tabular}{  c  c  c  }
\\ 
\hline \hline
Flash Figure    & $\Delta t $  & Bubble Wall Velocity  \\ 
\hline 
Figure \ref{091120012aZFit}     &       63 ns           &       -114 m/s         \\ 
Figure \ref{091120012bZFit}     &       47 ns           &       -136 m/s          \\ 
Figure \ref{typicalArFlashB}       &       92 ns           &       -91 m/s          \\ 
Figure \ref{typicalArFlash}          &       170 ns         &       -62 m/s           \\ 
Figure \ref{typicalKrFlash}          &       229 ns         &       -51 m/s           \\ 
Figure \ref{lateXenon}                 &       - 60 ns         &       47 m/s           \\ [1ex]
\hline
\end{tabular}    
\end{center}
\end{table}


\section*{Figure Captions}

\begin{figure}
\begin{center}
\includegraphics[scale=0.6]{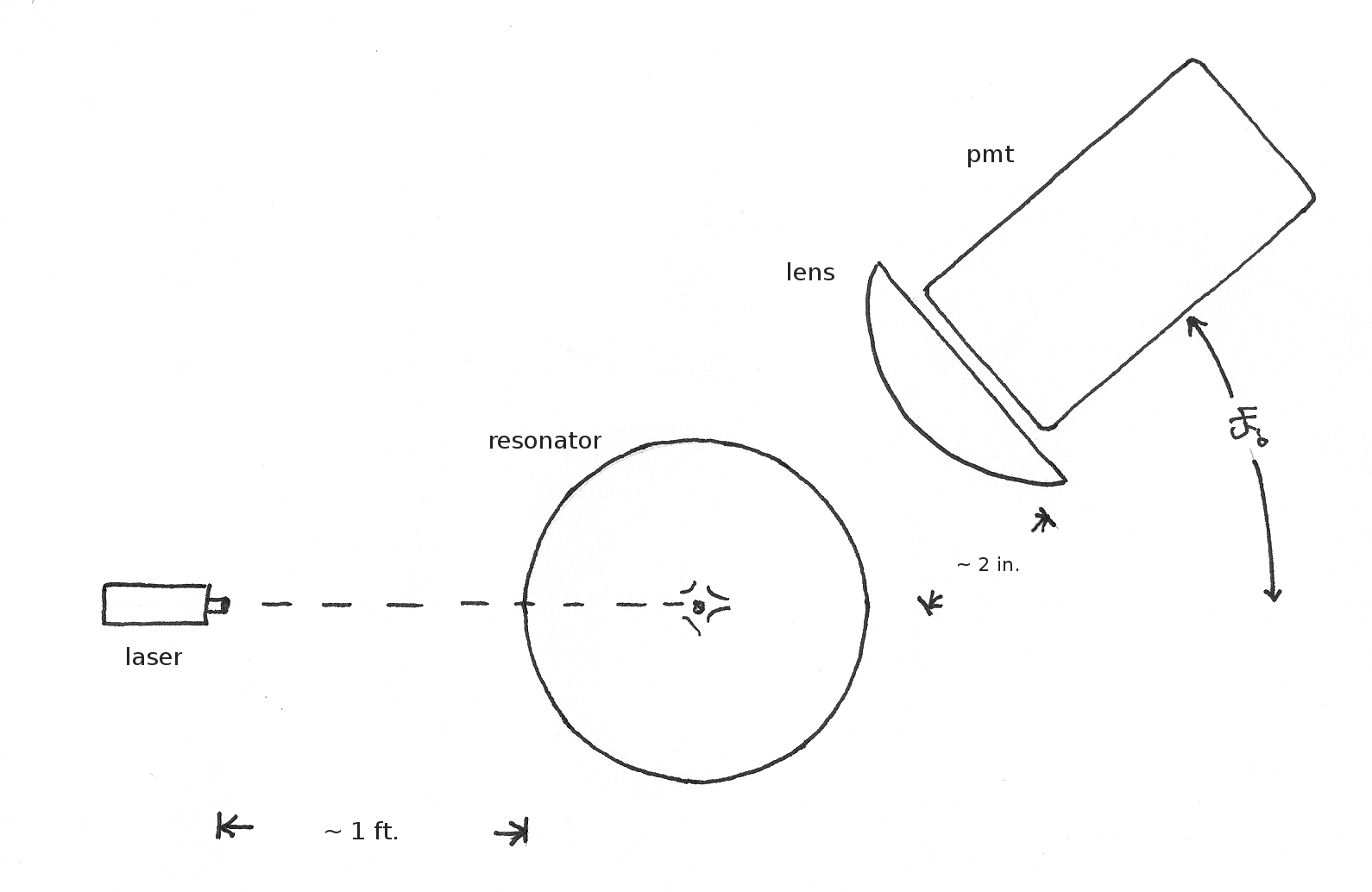}
\caption{\label{laserScatteringOverhead} An overhead schematic of the Flash Timing Experiment, showing approximate angles and distances.}
\end{center}
\end{figure}

\noindent {\bf  Figure \ref{laserScatteringOverhead}.}  An overhead schematic of the Flash Timing Experiment, showing approximate angles and distances.

\begin{figure}[h!tp]
\begin{center}
\includegraphics[scale=0.8]{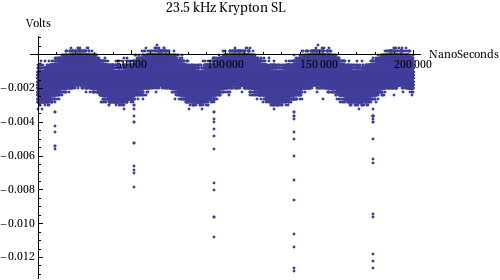}
\caption{\label{KryptonFlashToFlash3}  A scope trace showing five sequential SL flashes from a sonoluminescing krypton bubble.}
\end{center}
\end{figure}

\noindent {\bf Figure \ref{KryptonFlashToFlash3}.}  A scope trace showing five sequential SL flashes from a sonoluminescing krypton bubble.

\begin{figure}[h!tp]
\begin{center}
\includegraphics[scale=0.8]{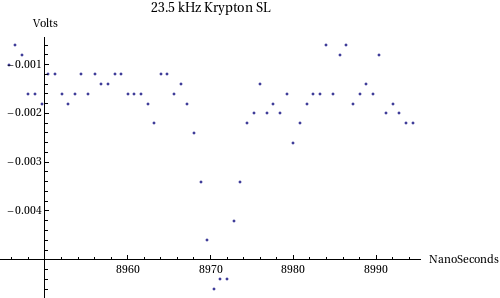}
\caption{\label{KryptonFlashToFlashA}  A zoom-in on the first flash from Figure \ref{KryptonFlashToFlash3}.  The peak of this flash is at 8970.4 ns.}
\end{center}
\end{figure}

\noindent {\bf Figure \ref{KryptonFlashToFlashA}.}  A zoom-in on the first flash from Figure \ref{KryptonFlashToFlash3}.  The peak of this flash is at 8970.4 ns.





\begin{figure}[h!tp]
\begin{center}
\includegraphics[scale=0.8]{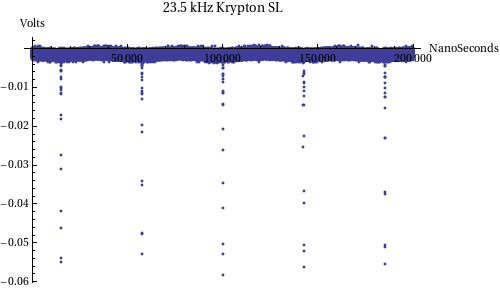}
\caption{\label{KryptonFlashToFlashZero}  Another scope trace showing five sequential SL flashes from a sonoluminescing krypton bubble.}
\end{center}
\end{figure}

\noindent {\bf Figure \ref{KryptonFlashToFlashZero}.}  Another scope trace showing five sequential SL flashes from a sonoluminescing krypton bubble.

\begin{sidewaysfigure}
\centering
\includegraphics[scale=0.8]{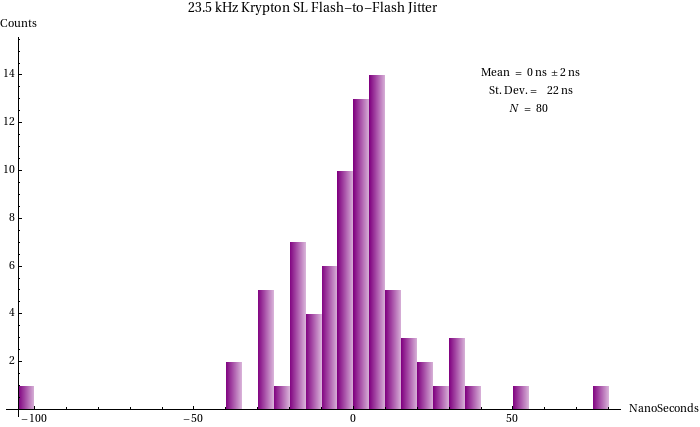}
\caption{\label{KryptonFlashToFlashHistogram}  A histogram of 80 flash-to-flash jitter measurements made from 100 krypton flashes.}
\end{sidewaysfigure}

\noindent {\bf Figure \ref{KryptonFlashToFlashHistogram}.}  A histogram of 80 flash-to-flash jitter measurements made from 100 krypton flashes.

\begin{sidewaysfigure}
\centering
\includegraphics[scale=0.60]{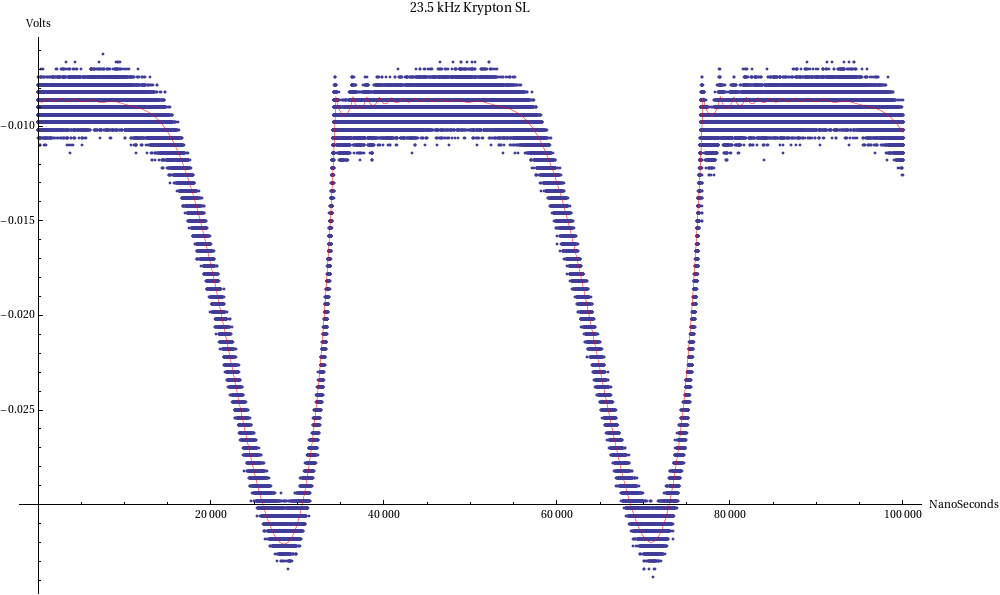}
\caption{\label{fig:091120012ManualFit} A scope trace of a sonoluminescing krypton bubble with fit to the Rayleigh-Plesset-Keller model.  This is from the 2009 dataset.}
\end{sidewaysfigure}

\noindent {\bf Figure \ref{fig:091120012ManualFit}.} A scope trace of a sonoluminescing krypton bubble with fit to the Rayleigh-Plesset-Keller model.  This is from the 2009 dataset.

\begin{sidewaysfigure}
\centering
\includegraphics[scale=0.60]{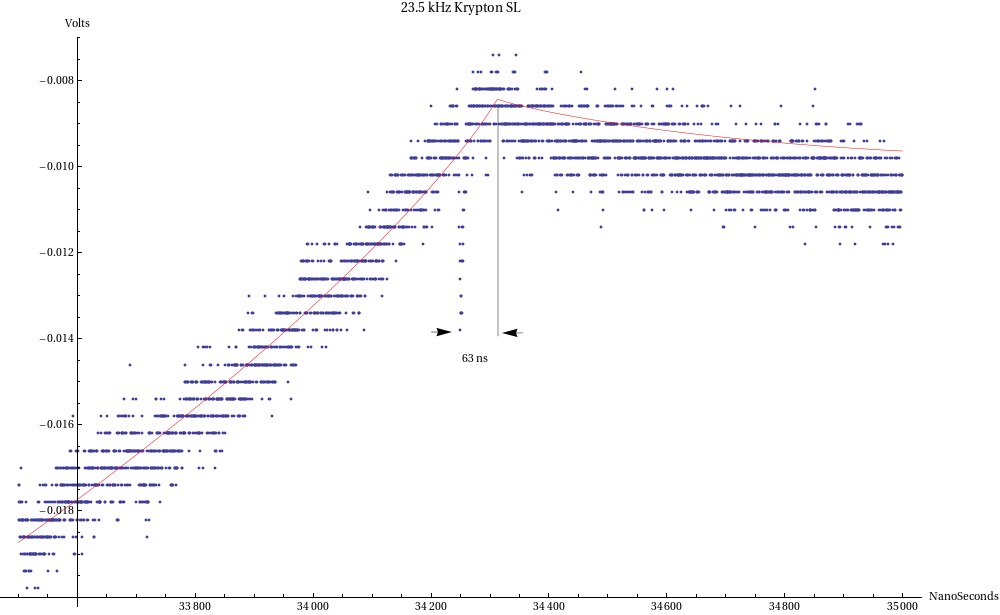}
\caption{\label{091120012aZFit} A zoom-in near the first minimum radius of the data from Figure \ref{fig:091120012ManualFit}, showing the flash occuring about 63 ns before the minimum radius.  The minimum radius time as determined from the fit is in approximate agreement with that determined by an ``eyeball'' reckoning.}
\end{sidewaysfigure}

\noindent {\bf Figure \ref{091120012aZFit}.} A zoom-in near the first minimum radius of the data from Figure \ref{fig:091120012ManualFit}, showing the flash occuring about 63 ns before the minimum radius.  The minimum radius time as determined from the fit is in approximate agreement with that determined by an ``eyeball'' reckoning.

\begin{sidewaysfigure}
\centering
\includegraphics[scale=0.60]{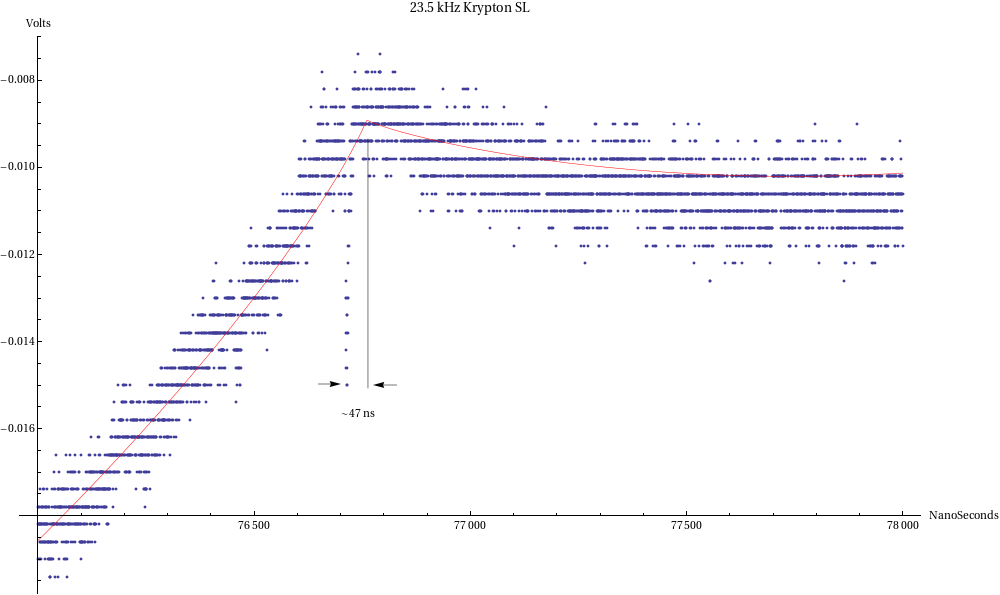}
\caption{\label{091120012bZFit} A zoom-in near the second minimum radius of the data from Figure \ref{fig:091120012ManualFit}, showing a flash about 47 ns before the minimum radius.}
\end{sidewaysfigure}

\noindent {\bf Figure \ref{091120012bZFit}.} A zoom-in near the second minimum radius of the data from Figure \ref{fig:091120012ManualFit}, showing a flash about 47 ns before the minimum radius.

\begin{sidewaysfigure}
\centering
\includegraphics[scale=0.5]{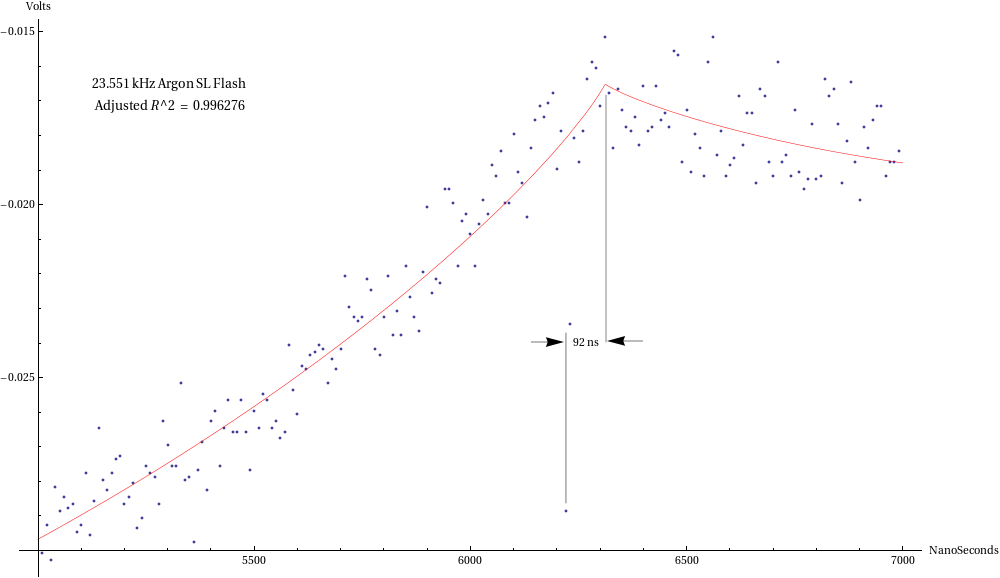}
\caption{\label{typicalArFlashB}A trace of a sonoluminescent argon bubble along with a three parameter fit to the RPK equation, with $R_0$ fixed at $9.5\,\mu m$, $P_d$ fixed at $1.4\,\text{atm}$ and $\alpha$, $V_{\text{bkg}}$ and $t_{\text{offset}}$ allowed to vary.  The results of the fit indicate that the flash is approximately $92\,\text{ns}$ before the minimum radius.}
\end{sidewaysfigure}

\noindent {\bf Figure \ref{typicalArFlashB}.} A trace of a sonoluminescent argon bubble along with a three parameter fit to the RPK equation, with $R_0$ fixed at $9.5\,\mu m$, $P_d$ fixed at $1.4\,\text{atm}$ and $\alpha$, $V_{\text{bkg}}$ and $t_{\text{offset}}$ allowed to vary.  The results of the fit indicate that the flash is approximately $92\,\text{ns}$ before the minimum radius.

\begin{sidewaysfigure}
\centering
\includegraphics[scale=0.4]{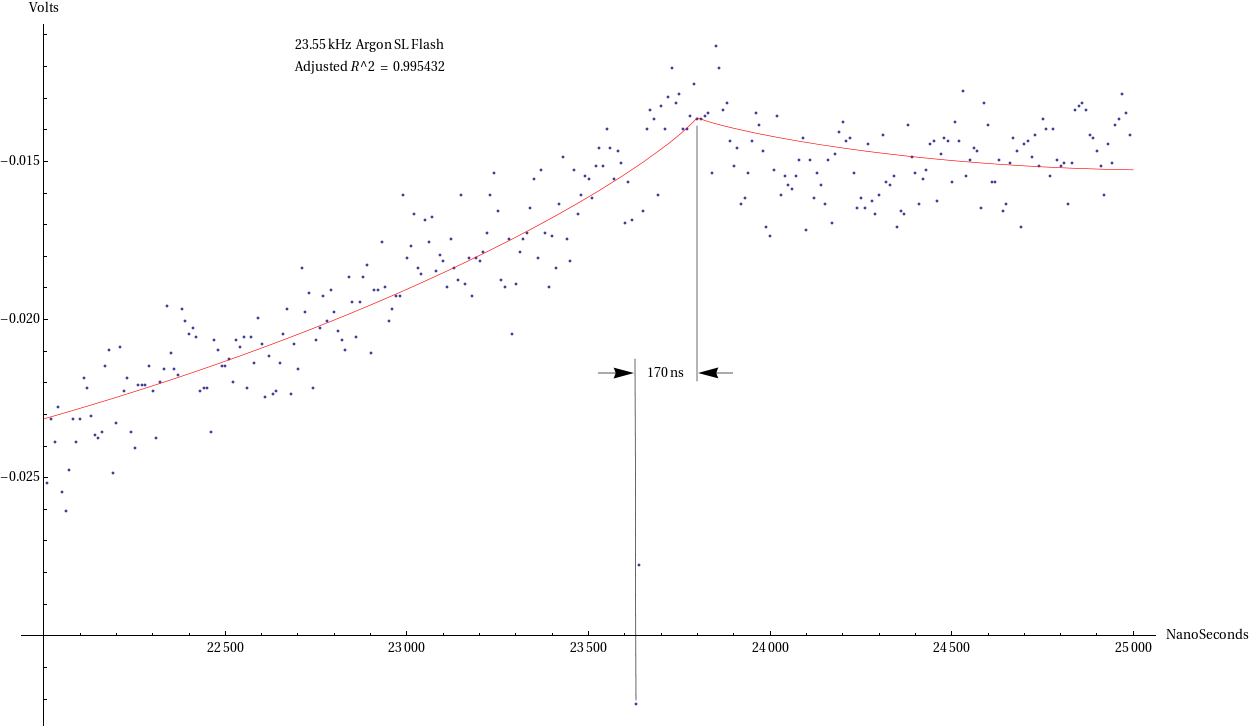}
\caption{\label{typicalArFlash}A trace of a sonoluminescent argon bubble along with a three parameter fit to the RPK equation, with $R_0$ fixed at $9.5\,\mu m$, $P_d$ fixed at $1.4\,\text{atm}$ and $\alpha$, $V_{\text{bkg}}$ and $t_{\text{offset}}$ allowed to vary.  The results of the fit indicate that the flash is approximately $170\,\text{ns}$ before the minimum radius.}
\end{sidewaysfigure}

\noindent {\bf Figure \ref{typicalArFlash}.}  A trace of a sonoluminescent argon bubble along with a three parameter fit to the RPK equation, with $R_0$ fixed at $9.5\,\mu m$, $P_d$ fixed at $1.4\,\text{atm}$ and $\alpha$, $V_{\text{bkg}}$ and $t_{\text{offset}}$ allowed to vary.  The results of the fit indicate that the flash is approximately $170\,\text{ns}$ before the minimum radius.

\begin{sidewaysfigure}
\centering
\includegraphics[scale=0.4]{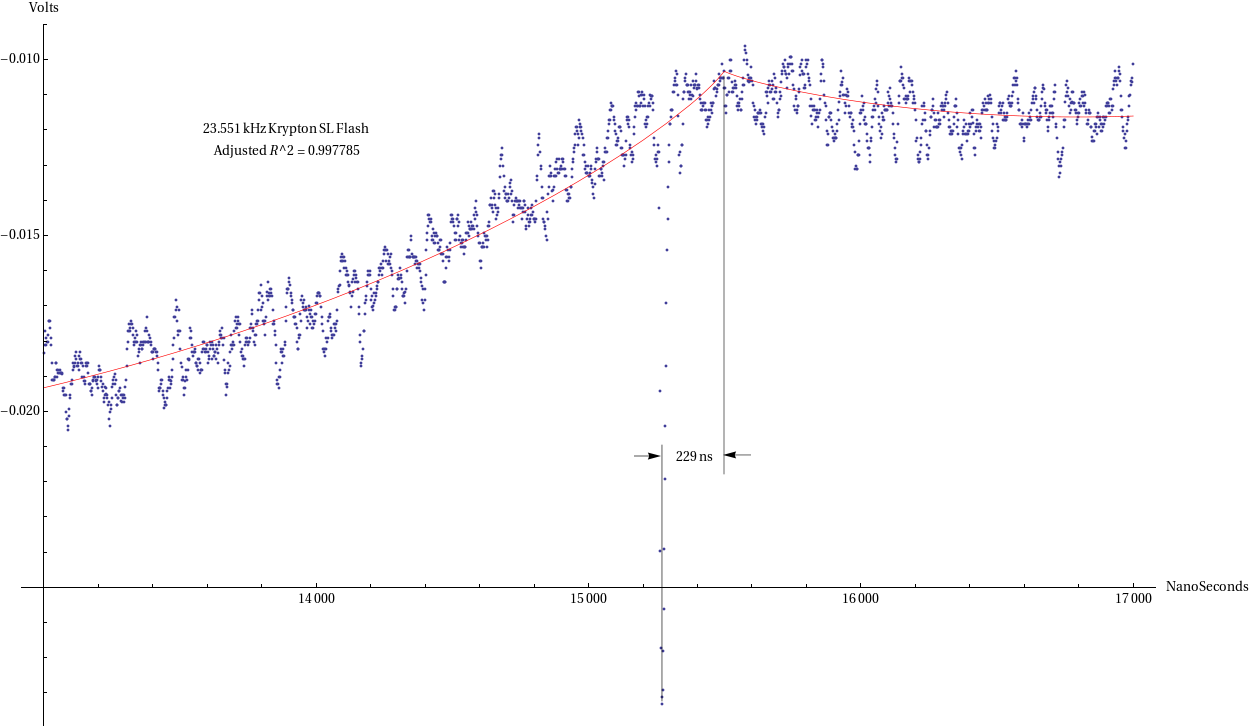}
\caption{\label{typicalKrFlash}A trace of a sonoluminescent krypton bubble along with a three parameter fit to the RPK equation, with $R_0$ fixed at $9.5\, \mu m$, $P_d$ fixed at $1.4 \, \text{atm}$ and $\alpha$, $V_{\text{bkg}}$ and $t_{\text{offset}}$ allowed to vary.  The results of the fit indicate that the flash is approximately $229\,\text{ns}$ before the minimum radius.}
\end{sidewaysfigure}

\noindent {\bf Figure \ref{typicalKrFlash}.}  A trace of a sonoluminescent krypton bubble along with a three parameter fit to the RPK equation, with $R_0$ fixed at $9.5\, \mu m$, $P_d$ fixed at $1.4 \, \text{atm}$ and $\alpha$, $V_{\text{bkg}}$ and $t_{\text{offset}}$ allowed to vary.  The results of the fit indicate that the flash is approximately $229\,\text{ns}$ before the minimum radius.

\begin{sidewaysfigure}
\centering
\includegraphics[scale=0.6]{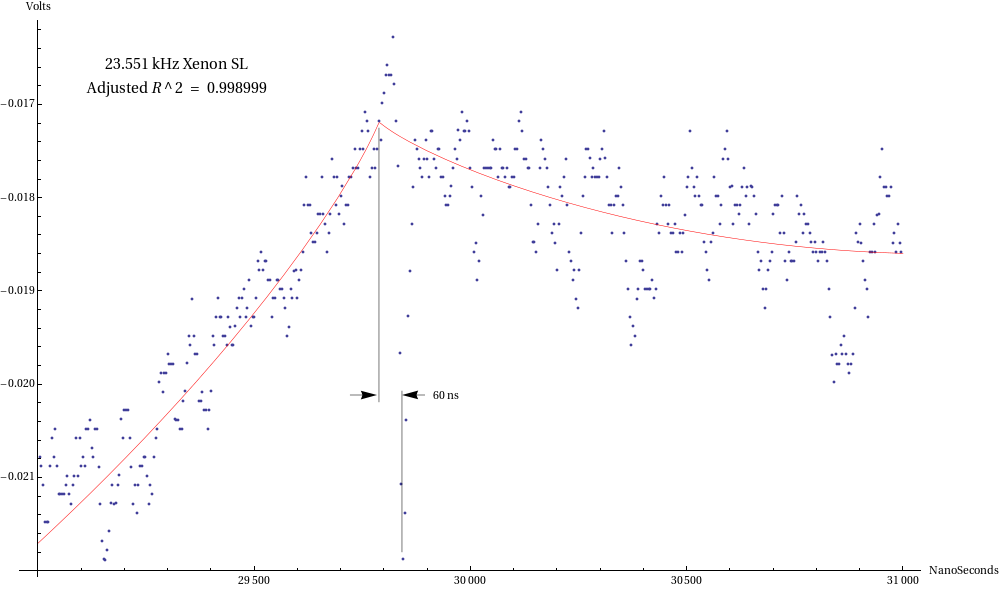}
\caption{\label{lateXenon}  A xenon SL flash from the 2008 dataset .  These data show the peak of the flash occuring $60\, \text{ns}$ after the minimum radius.}
\end{sidewaysfigure}

\noindent {\bf Figure \ref{lateXenon}.}  A xenon SL flash from the 2008 dataset .  These data show the peak of the flash occuring $60\, \text{ns}$ after the minimum radius.


\clearpage

\clearpage

\clearpage






\clearpage

\clearpage

\clearpage

\clearpage

\clearpage

\clearpage

\clearpage

\clearpage

\clearpage

\end{document}